\documentclass[twocolumn,pre,floats,aps,amsmath,amssymb]{revtex4}
\usepackage{graphicx,amsmath,subfigure}
\usepackage{bm}
\usepackage{natbib}
\begin{document}

\title{Hardware Random number Generator for cryptography}
\author{Ram Soorat\footnote{corresponding author email:rsoorat@gmail.com}, Madhuri K. and  Ashok Vudayagiri}
\affiliation{School of Physics, University of Hyderabad, Hyderabad 500046, India} 
\date{\today}
\begin{abstract}
One of the key requirement of many  schemes is that of random numbers. Sequence of random numbers are used at several stages of a standard cryptographic protocol. A simple example is of a Vernam cipher, where a string of random numbers is added to massage string to generate the encrypted code. It is represented as  $C=M \oplus K $  where $M$ is the message, $K$ is the key and $C$ is the ciphertext. It has been mathematically shown that this simple scheme is unbreakable is key K as long as  M and is used only once. For a good cryptosystem, the security of the cryptosystem is  not be based on keeping the algorithm secret but solely on keeping the key secret.  The quality and unpredictability of secret data is critical to securing communication by modern cryptographic techniques. Generation of such data for cryptographic purposes typically requires an unpredictable physical source of random data. In this manuscript, we present studies of  three different methods for producing random number. We have tested them  by studying its frequency, correlation as well as using  the test suit from NIST.
\end{abstract}
\maketitle
\section{Introduction}
 A random process is a repeating process in which output is difficult to find a describable deterministic pattern. The term randomness is quite often used in statistics to signify well defined statistical properties, such as correlation. Saying that a variable is random means that the variable follows a given probability distribution. In such a definition, random is different from arbitrary. A good RNG should work efficiently, which means it should be able to produce a large amount of random numbers in a short period of time. Random numbers are widely used in many applications, such as cryptography \cite{bb84,crypto},  spread-spectrum  communications  \cite{pickholtz},  Monte  Carlo  numerical  simulations  \cite{metro},  and  ranging \cite{naza}  statistical analysis, numerical simulations, information security, stochastic simulation, stream ciphers, ranging  signal in radar system, controlling signal in remote control, encryption codes or keys in digital  communication,  address  codes  and  spread  spectrum  codes  in  code  division  multiple  access  (CDMA) and many others. So simulations of random numbers are crucial. amounts of random numbers are necessary and thus fast RNGs are required. They are also used in the statistics to solve problems in many fields such as nuclear medicine, finance and computer graphics.
 
There are in general two types of generators for producing random sequences: true random number generators (TRNGs) and pseudo random number generators (PRNGs).   PRNGs need some input called seeds, along with some deterministic algorithms to generate multiple pseudo random  numbers. They are usually faster than TRNGs and are preferable when a lot of random-like numbers are required. Since traditional random numbers generated by algorithms are essentially pseudo-random, they have potential danger in security-related fields like quantum key distribution. TRNGs make use of non-deterministic sources along with some post-processing functions for generating randomness. Such sources include physical phenomena such as thermal noise, atmospheric noise, radioactive decay and even coin tossing. such as electrical  noises \cite{petrie}, frequency jitters in electrical oscillators \cite{bucci} and chaotic circuits \cite{stoj,stoj1}, which can  produce  unpredictable  random  numbers  of  high  quality  yet  much  lower  rates  than  PRNGs  because of the narrow bandwidth of these physical entropy sources. In addition a number of documents exist which  provide general advice on using and choosing random number sources \cite{callas,tim,crypto1,klaus}. Further discussions on the nature of randomness, pseudo random number generators (PRNG’s), and cryptographic randomness are available from a number of sources \cite{donald,paul,crypto2}.
 
A true Random numbers is the base of many cryptographical applications like QKD, especially to generate keys that cannot be penetrated by hackers or other attackers it is important that the random numbers used is unpredictable. The BB84 protocol makes use of polarization states of single photons to map the bits 0, 1 of the encryption key, in two mutually unbiased basis. This results in pulses containing single photons, each randomly this means that the RNG has to perform better than the Pseudo random number generators (PRNG) available on the computer, but also should be compact and easy to integrate into the prototype QKD device. This requires a controller that generates four random states and its deterministic critically endangers the security of the entire protocol. For most applications it is desirable to have fast random number generators (RNGs) that produce numbers that are as random as possible.

\section{Test for Randomness}
There are different types of  statistical tests  that can be applied to a sequence to attempt to compare and evaluate the sequence to a truly random sequence. Random sequence can be characterized and described in terms of probability.  In addition, the results of statistical testing must be interpreted with some care and caution to avoid incorrect conclusions about a specific generator. We use  a set tests designed and implemented by national institute of standard and technology (NIST) test suite is a statistical package consisting of 15 tests \cite{andrew}.
This package will address the problem of evaluating (P)RNGs for randomness. It will be useful in
\begin{itemize}
\item Identifying (P)RNG's which produce weak (or patterned) binary sequences,
\item Designing new (P)RNG's,
\item Verifying that the implementations of (P)RNG's are correct,
\item Studying (P)RNG's described in standards, and
\item Investigating the degree of randomness by currently used (P)RNG's.
\end{itemize}
\subsection{P-value}
The probability (under the null hypothesis of randomness) that the chosen test statistic will assume values that are equal to or worse than the observed test statistic value when considering the null hypothesis. The P-value is frequently called the "tail probability".

If a P-value for a given run equals to 1, then the sequence is decided to be  perfectly random.  A P-value of zero indicates a completely non-random sequence. In addition a significance level ($\alpha$) can be chosen such that  if $P-value \geq\alpha$, then the sequence can be considered to be almost random.  If $P-value < \alpha$,  the sequence appears to be non-random. The parameter $\alpha$ denotes the probability of the Type I error and typical values for $\alpha$ is chosen in the range [0.001, 0.01].

 An $\alpha$ of 0.001 indicates that one would expect one sequence in 1000 sequences to be rejected by the test if the sequence was random. For a $P-value \geq 0.001$, a sequence would be considered to be random with a confidence of 99.9\%. For a $P-value < 0.001$, a sequence would be considered to be non-random with a confidence of 99.9\%.

\section{ Generation of  Random Numbers } 
A Random numbers are important requirement for both classical encryption as well as QKD. Several attempts were made for random number generator, which would be used for the entire QKD scheme. We present here our studies of three different RNGs, 
\begin{itemize}
\item A hardware generator built around Kuuselas chaos circuit
\item Dark Counts from an Avalanche Photo Diode
\item Pseudo Random code from internal modules of  LabVIEW
\end{itemize}
In all these methods we show the efficiency of random number generator by analyzing their statistics.
\section{Chaos Based Hardware Random Number Generator}
The circuit is based on a  chaos based hardware random number generator which consists of an inductor and capacitance diode, called the Varactor.
\begin{figure}[h!]
\centering
\includegraphics[height=6cm,width=8cm]{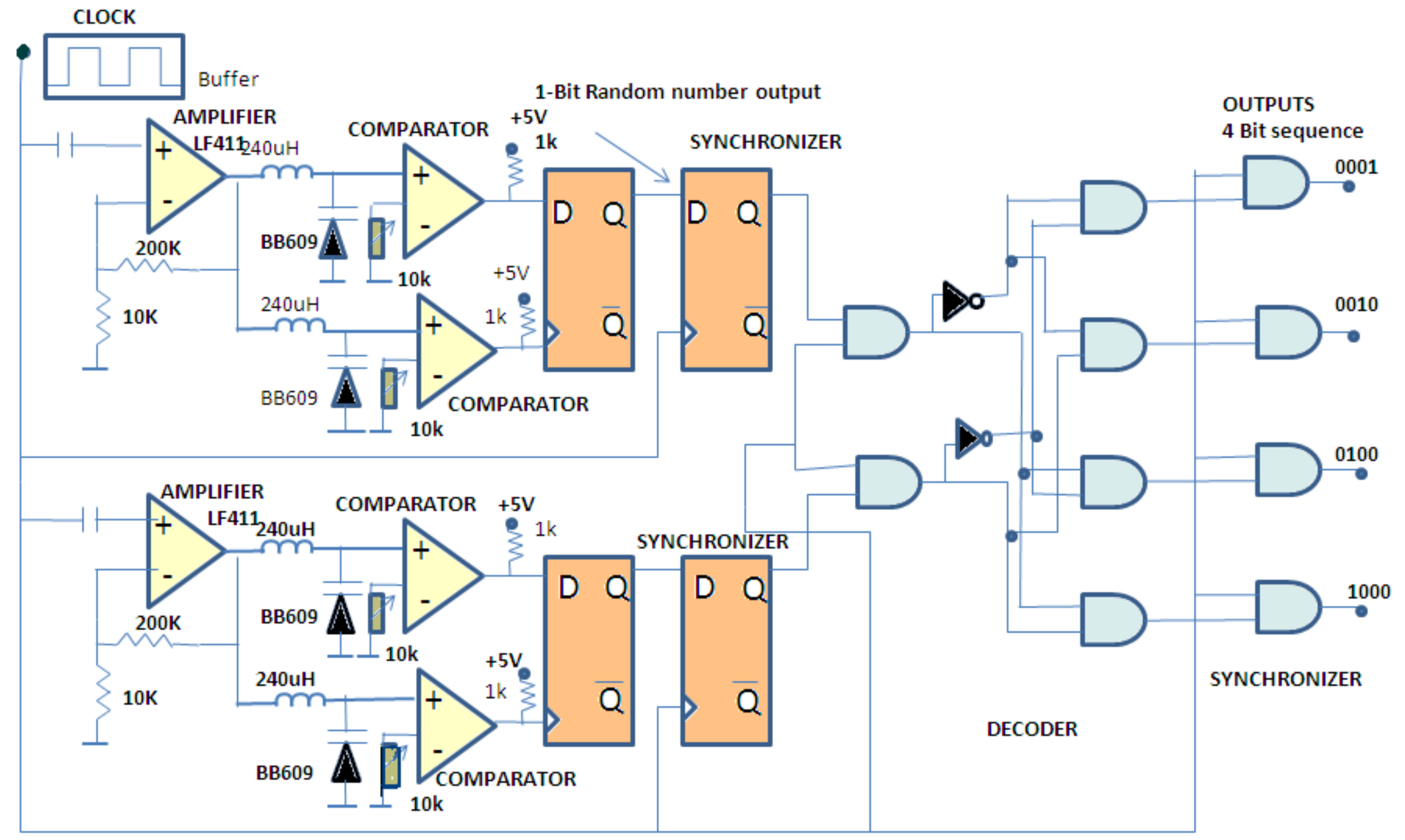}
\caption{Circuit Diagram}
\label{kussela}
\end{figure}
The figure \ref{kussela} is a simple LCR circuit built around a varactor diode. Due to its voltage dependent capacitance act as non linear element thus providing a chaotic oscillation.

Generally a hardware random number generator is based on sampling noise sources such as thermal noise or reverse based diode. Different circuits are described in  but these methods are difficult since there amplitudes are usually small and often masked by deterministic disturbances so the other alternative is to use a chaotic oscillator for pseudo random generator due to its unpredictable behavior and relatively simple. We have built a chaotic circuit based RNG (CCRNG), based on an earlier design by T.Kuusela \cite{kuusela}, which in turn is built around the chaos generator of Matsumoto et. al \cite{matsumoto}.  The nonlinear element here  is the varactor,  whose capacitance varies as a function of voltage across it.  The capacitance is varied as $C (V) =C/ (1+V/\theta)\gamma$,  where V is the voltage across the diode. If the circuit parameters and the external drive are suitably chosen, the system exhibits period doubling and chaotic behaviour  \cite{matsumoto}.

Our circuit for  chaotic clock Generator (CCG)  is as shown in \ref{kussela}. It has an external clock signal (A square wave of 500-600 KHz is used as a clock in this case). Our CCG includes two identical chaotic oscillators so as to  generate a sequence of four bit structure in a random fashion, viz.,0001, 0010, 0100 and 1000. The heart of the unit is a pair of inductor 240$/mu$H and a varactor diode BB609.  An amplifier is used to raise the signal level so that if the frequency and the amplitude of the driving clock is properly chosen the circuit goes to chaotic state and a reliable operation is guaranteed even the case of large tolerance. The fast operational amplifier, LF411,  with a large bandwidth is used as an amplifier. The voltage across the capacitor diode is a random signal when the circuit exhibits period doubling and chaotic behaviour. A Comparator (LM 311 with response time of 200ns) is used to convert analog signal into a digital signal. Even though the output of the comparators are quite random to generate it does not generate all the possible bit sequences. To overcome this, we use two comparators for each CCRNG one of them takes bit sample from the other this is done by a simple D-flip flop(1st D-flip flop). To synchronize the output bit sequences with the driven clock one more D-flip flop is used and a AND gate is used to avoid continues 1’s and 0’s. A decoder is used to generate four bits structures using two CCRNG’s.

\begin{figure}[h!]
\centering
\includegraphics[height=8cm,width=8cm]{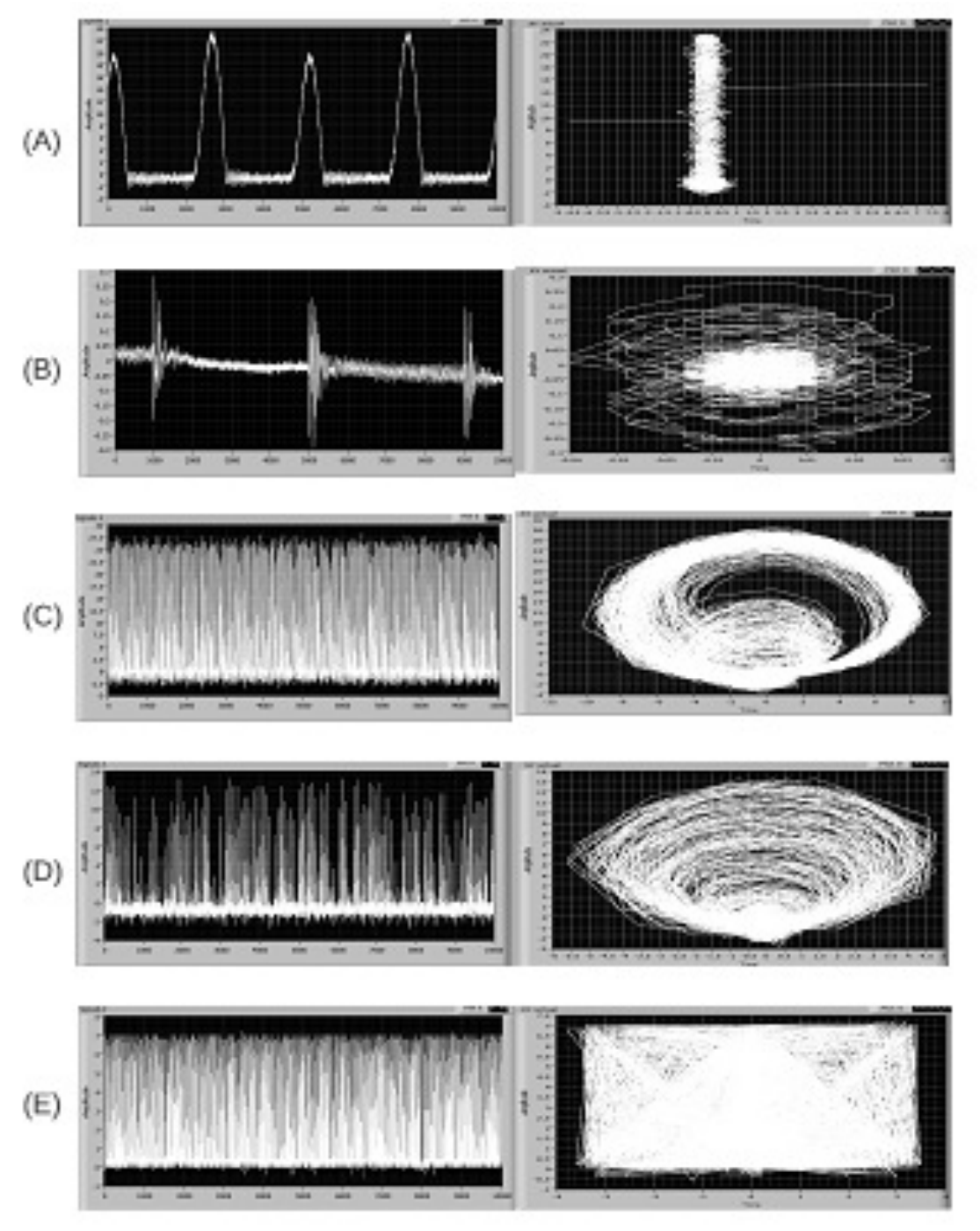}
\caption{Analog signal output of the Chaos
Clock Generator at different frequencies and
corresponding Phase plots}
\label{hphase}
\end{figure}

Using this circuit we generated several sets of random sequence of 0's and 1's and tested its behaviour. At first we tapped the voltage at the output of the chaotic circuit, at the edge of the varactor. This data stored onto the computer as a function of time and  its variation $\frac{dv}{dt}$ is computed. Plotting $\frac{dv}{dt}$ against V gives a phase plot.  Figure \ref{hphase} shows behavior of the generated signal. The plot on left side shows the raw data of voltage  v/s time while those on right side shows the phase  plot $\frac{dV}{dt}$ v/s V. As the frequency of the input clock circuit  is changed the circuit goes from a monotonic oscillator (figure A), to a bifurcation (figure C and D) and finally to chaos (figure E). Output becomes chaotic when clock frequency is around  650 kHz (figure E).

The comparator converts this signal into a sequence of 0's and 1's.  If the output of chaotic circuit is above a set threshold value, the comparator gives a one or else a zero. The output pulse is synchronized to the clock pulse. The pulses   are recorded on to the computer and further analyzed. these data are obtained at the clock speed of about 650 kHz, since this is the region when chaos circuit is giving a proper chaotic output.

\begin{figure} [!h]
\centering
\includegraphics[height=4cm,width=4cm]{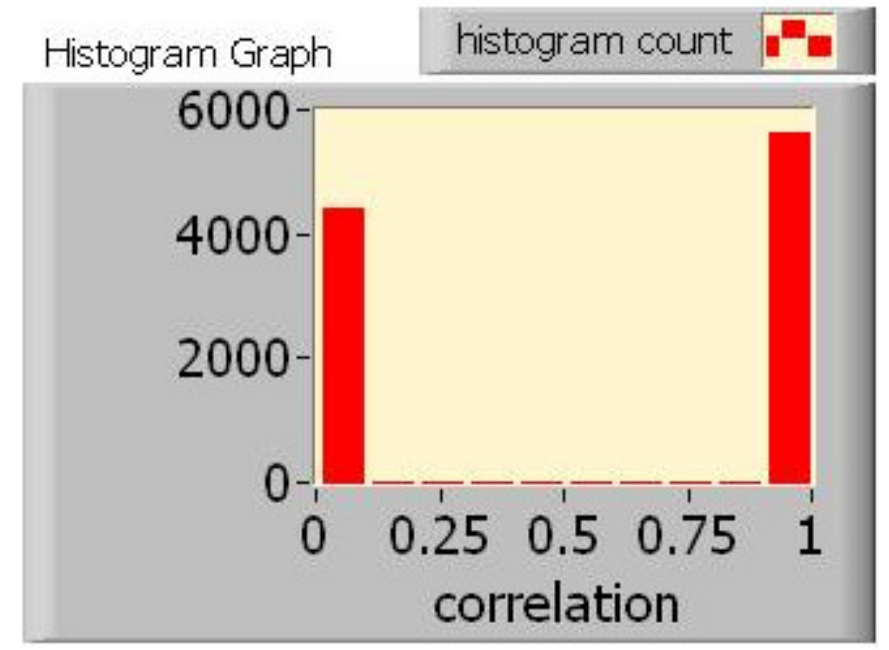} 
\includegraphics[height=4cm,width=4cm]{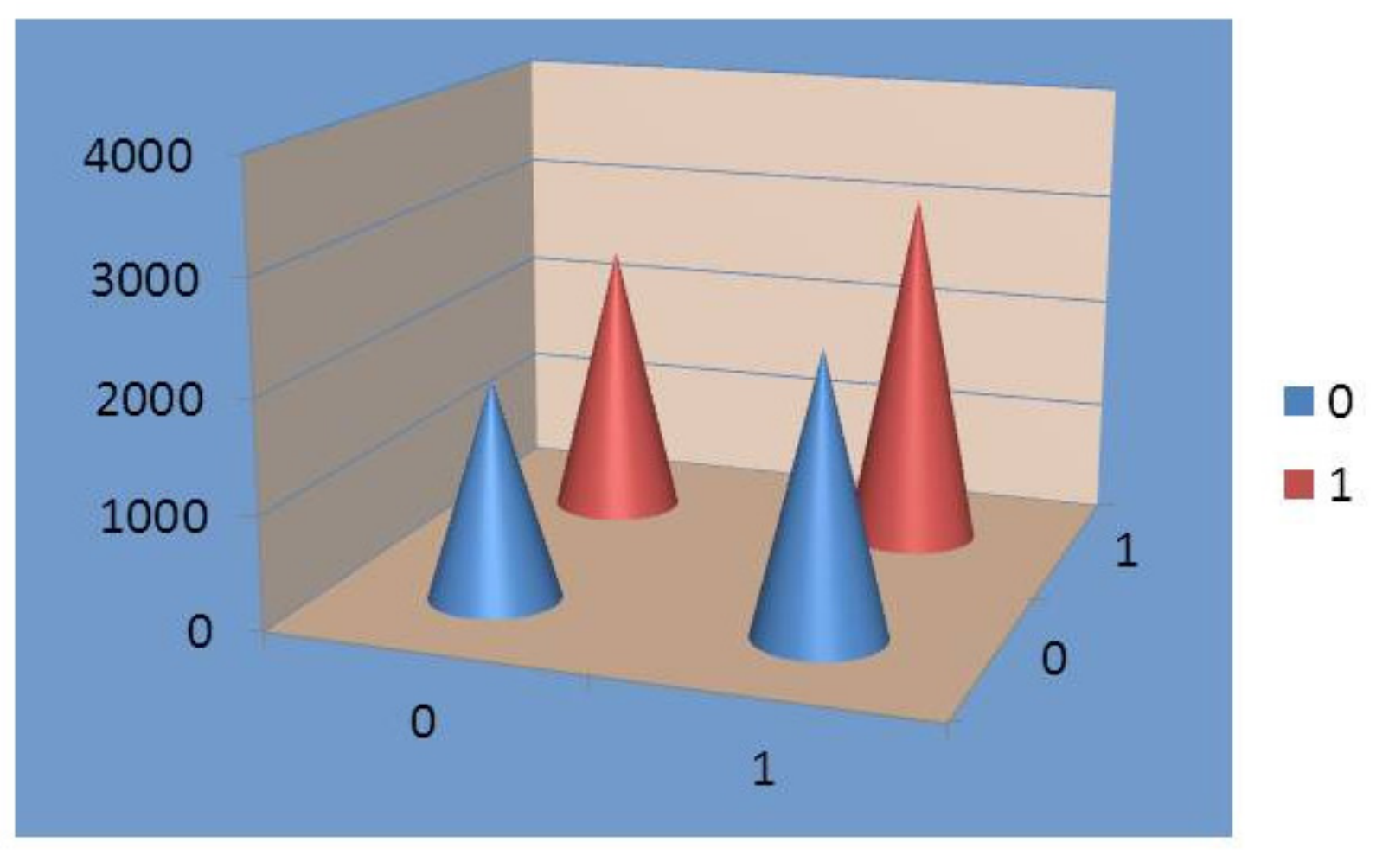}
\caption{Histogram distribution of 0 and 1 (b) bit correlation}
\label{hhisto}
\end{figure}

Figure \ref{hhisto} shows two types of distribution. The total number of 1's and 0's are shown in left side . This run is  asymmetric since there is more 1's than 0. This happen only for a few runs and can easily be corrected by changing the threshold value. 

The graph on right, showing cross correlations is more important.  It tells the probability of getting zero following 1, getting a 1 after 1 and  0 after 0 and 1 after 0. the graph shows a slight higher value of occurrence of `1,1' sequence but this is due to asymmetry of our circuit, which causes  more occurrences of 1's as opposed to zero. More importantl, the graph clearly shows same frequency for 10 and 01. This means that the system does not have preference for 1 over 0, and a sequence of '0,1' is as probably as that of '1,0'.

\begin{figure} [h!]
\centering
\includegraphics[height=5cm,width=8cm]{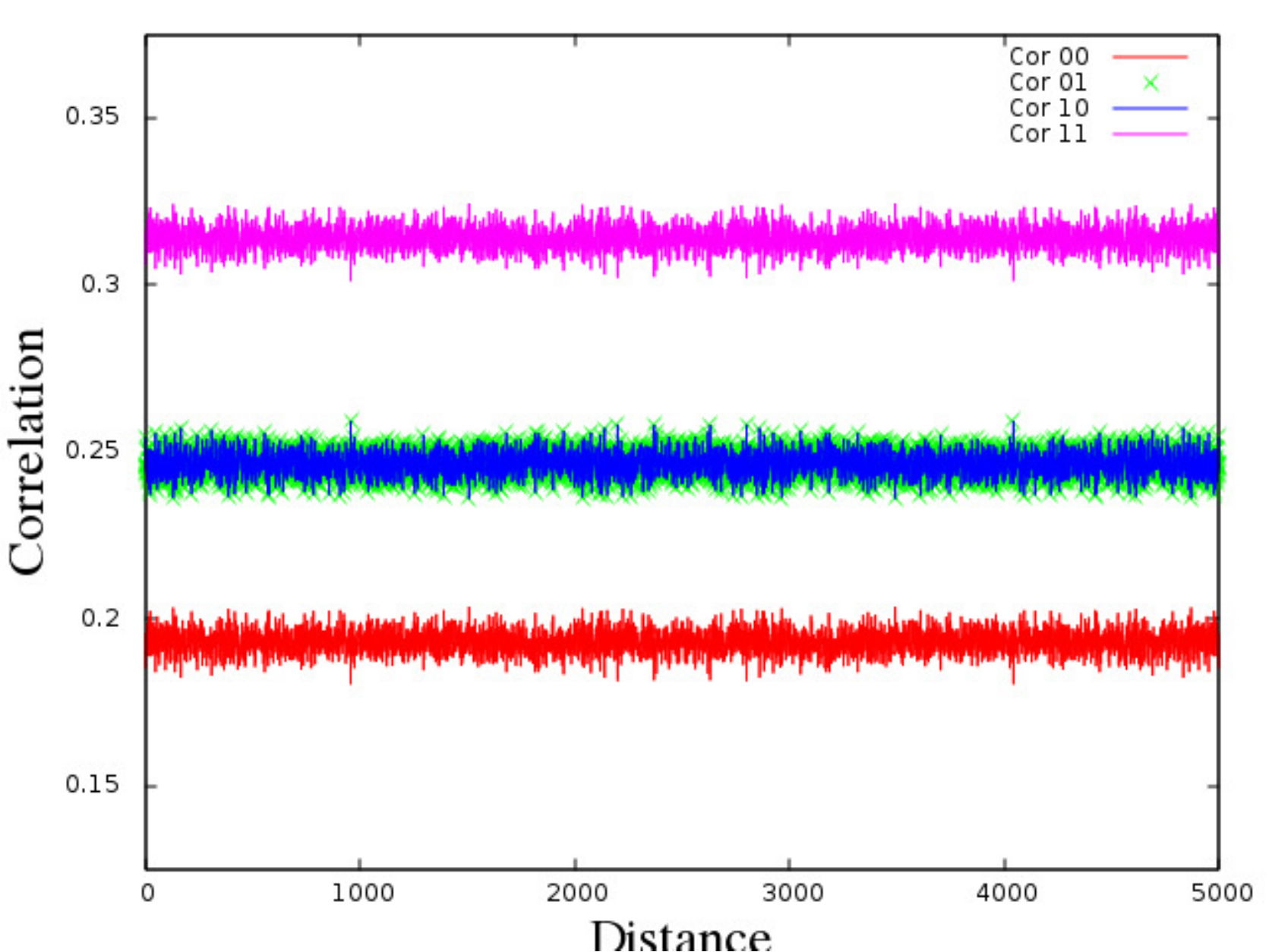}
\caption{Successive  bit correlation }
\label{hbitcor}
\end{figure}

In addition, we studied the bit correlation  of the data at different levels that is $c_{ij}^{00}=< P_i (0)P_j (0)>$ where $P_i (0)$ and $P_j (0)$ are the probability of finding 0 at $i^{th}$ position and $j^{th}$ position respectively. Extending this calculation, we can write $c_{ij}^{kl}=< P_i (k)P_j (l)>$ for k=1,0 and l=1,0. For an ideal case all this should be equal and equal to a value 0.25. The graphs are shown in figure \ref{hbitcor}, where `Distance' represents the gap between `$i$' and `$j$'. 

However, figure \ref{hbitcor} shows that $c_{ij}^{11}=0.35$ and $c_{ij}^{00}=0.9$ where as $c_{ij}^{10}=c_{ij}^{01}=0.25$. This discrepancy in again due to the fact that the circuit is slightly biased towards 1, due to the setting on the discriminator. On the other hand  the correlation between 01 and 10 are both equal to 0.25, for all values of $i$ and $j$.  indicating the circuit is a near perfect coin toss system. 


We also processed the sequence of 0's and 1's through the NIST test suite. 
The P values for all the 15 tests are given in table \ref{nist_kuusela}
\begin{table} [h!]
\center
\begin{tabular}{|c|c|}
\hline
\textbf{ Statistical test} &\textbf{P-value }\\
\hline
 Frequency & 0.000199 \\
 BlockFrequency &0.000003 \\
CumulativeSums & 0.000439\\
CumulativeSums& 0.017912\\
 Runs & 0.350485\\
LongestRun &  0.066882 \\
 Rank & 0\\
 FFT & 0.017912 \\
 NonOverlappingTemplate & 0.004301 \\
 NonOverlappingTemplate & 0.002043 \\
 NonOverlappingTemplate &  0.350485 \\
 NonOverlappingTemplate &0.035174\\
 \hline
\end{tabular}
\caption{RESULTS FOR THE UNIFORMITY OF P-VALUES AND THE PROPORTION OF PASSING SEQUENCES }
\label{nist_kuusela}
\end{table}

The P values are very low indicating that this is not a very reliable RNG, even though the bit sequence is not correlated. This could be due to the improper biasing of the system and needs to be further examined. 

\section{Random number using LabVIEW }
LabVIEW is a proprietary software by National Instruments Inc. USA, mainly designed to interface hardware components of an experimental setup to the computer. Since we extensively use this software to connect our data acquisition and processing, we decided to test the in-built RNG or LabVIEW. The in-built module of LabVIEW Produces a double precision, floating-point number between 0 and 1, exclusively. Documentation of NI declares that this module uses a multiple multiplicative congruential generators, using system clock as a seed. 

\begin{figure}[h!]
\centering
\includegraphics[height=6cm,width=8cm]{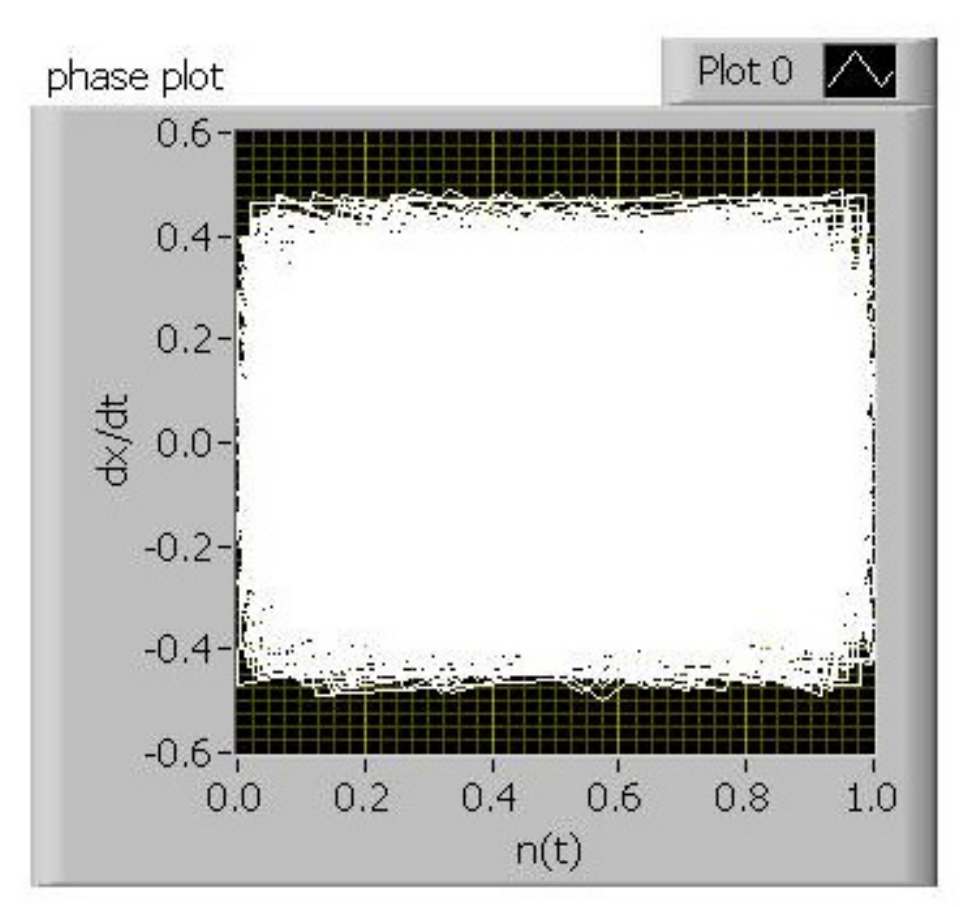}
\caption{Phase plots}
\label{lphase}
\end{figure}

The output of this module has a uniform distribution. At first we test  $\frac{dn}{dt}$ v/s n, as shown in  figure \ref{lphase}. In this case $n$ is the bit output, with values $0$ or $1$. The figure essentially shows the probability of getting a fixed value at time t and its dependence on previous value. The highly non-periodic phase plot, indicates a randomness. 

\begin{figure} [h!]
\centering
\includegraphics[height=4cm,width=4cm]{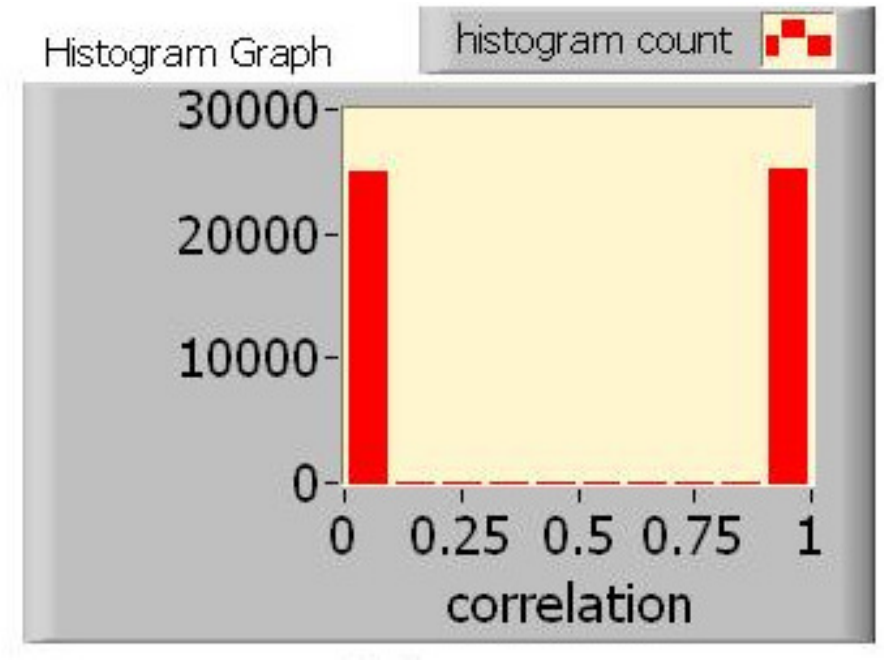}
\includegraphics[height=4cm,width=4cm]{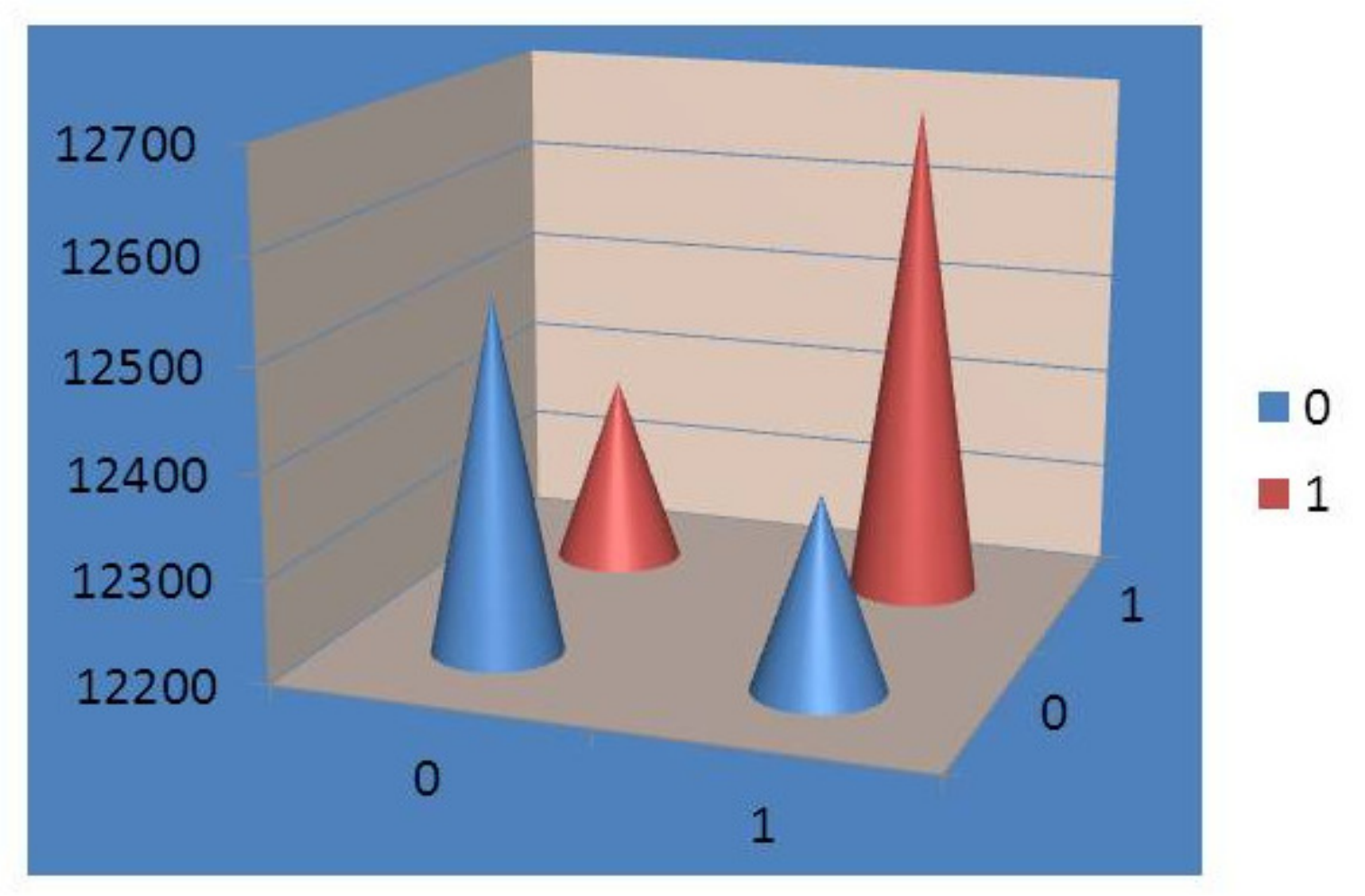}
\caption{Histogram distribution of 0 and 1 (b) bit correlation}
\label{lcor}
\end{figure}

The figure \ref{lcor}  is  histogram graph which  is a  visual representation of data that measures the number of incidents of 0 and 1 of a  sample set. left side  shown the distribution  of 1's and 0's. We  observe  same  frequency for 1's and 0's which is  true randomness behaviour. But the graph clearly shown same frequency for 10 and 01. This means that the system does not have any performance for 1 over 0.

\begin{figure} [!h]
\centering
\includegraphics[height=6cm,width=8cm]{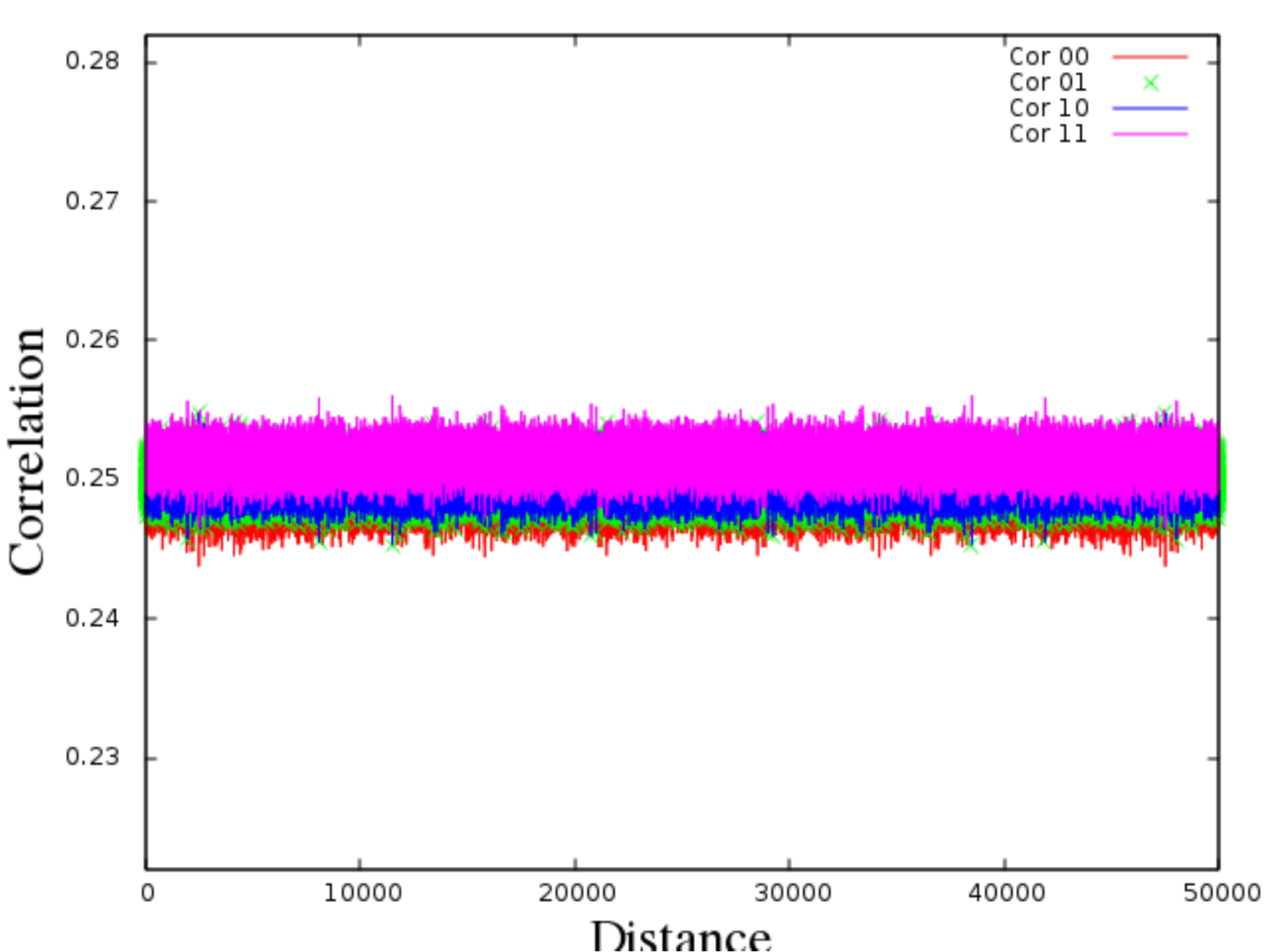}
\caption{Successive  bit correlation }
\label{lhcor}
\end{figure}

As in case of earlier, we also compute correlation between $i^{th}$ bit and $j^{th}$ bit. Except for a small bias, of about 0.3, for correlation for $i=1$ and $j=1$, remaining correlations, shown in figure \ref{lcor} are nearly 0.25, for all values of $i,j$. 

We process the NIST suite for this data as well and get a better result. The corresponding P values are shown in table \ref{l_nist}.

\begin{table}[!h]
\center
\begin{tabular}{|c|c|}
\hline
\textbf{ Statistical test} &\textbf{P-value }\\
\hline
 Frequency & 0.739918 \\
 BlockFrequency &0.350485 \\
CumulativeSums & 0.534146\\
CumulativeSums& 0.534146\\
 Runs & 0.911413\\
LongestRun & 0.739918 \\
 Rank & 0\\
 FFT & 0.066882 \\
 NonOverlappingTemplate & 0.122325 \\
 NonOverlappingTemplate & 0.035174\\
 NonOverlappingTemplate & 0.213309\\
 NonOverlappingTemplate & 0.066882\\
 NonOverlappingTemplate & 0.066882 \\
 NonOverlappingTemplate & 0.000954\\
 NonOverlappingTemplate &0.004301\\
 NonOverlappingTemplate & 0.017912\\
\hline
\end{tabular}
\caption{RESULTS FOR THE UNIFORMITY OF P-VALUES AND THE PROPORTION OF PASSING SEQUENCES }
\label{l_nist}

These P values, shown in table \ref{l_nist} are more than 0.5 for almost all the tests. It means its a good random that we can use for encryption.
\end{table}

\section{Conclusion}
Since software methods only offer a pseudo random number codes, need for other sources is important. We have therefore analyzed hardware based  random number generator. In this we have compare and analyze the randomness behavior between software and hardware random number generator. Although the correlation test gives very good result, the NIST test  shows it is yet a poor option, and that needs to be rectified. The result of correlation indicates that this has a promise and can be worked on. We show the in-built PRNG of LabVIEW shows a relatively better performance. Additional work is needed to tweak our CCRNG to show a better performance.


\begin{thebibliography}{00}
\bibitem{bb84} C.H. Bennett and G. Brassard In: Proc. IEEE Int. Conf. on Com-
puters, Systems, and Signal Processing at Bangalore, IEEE, New York
, p. 175.1984
\bibitem{crypto}  Security requirements for cryptographic modules. FIPS 140–2 (2001).  http://csrc.nist.gov/publications/fips/fips140-2/fips1402.pdf 
\bibitem{pickholtz}  R. L. Pickholtz, D. L. Schilling, and L. B. Milstein, “Theory of spread-spectrum communications-a tutorial,” IEEE Trans. Commun. 30(5), 855–884, 1982 
\bibitem{metro}  N. Metropolis and S. Ulam, “The Monte Carlo method,” J. Am. Stat. Assoc. 44(247), 335–341,1949 
\bibitem{naza}  M. Nazarathy, S. A. Newton, R. P. Giffard, D. S. Moberly, F. Sischka, W. R. Trutna, Jr., and S. Foster, “Realtime long range complementary correlation optical time domain reflectometer,” J. Lightwave Technol. 7(1), 24–38, 1989 
\bibitem{petrie} C. Petrie and J. Connelly, “A noise-based IC random number generator for applications in cryptography,” IEEE  Trans. Circ. Syst. I Fundam. Theory Appl. 47(5), 615–621, 2000 
\bibitem{bucci} M. Bucci, L. Germani, R. Luzzi, A. Trifiletti, and M. Varanonuovo, “A high-speed oscillator-based truly random  number source for cryptographic applications on a smart card IC,” IEEE Trans. Comput. 52(4), 403–409, 2003 
\bibitem{stoj} T. Stojanovski and L. Kocarev, “Chaos-based random number generators-Part I: analysis,” IEEE Trans. Circ. Syst. I Fundam. Theory Appl. 48(3), 281–288, 2001 
\bibitem{stoj1}  T. Stojanovski, J. Pihl, and L. Kocarev, “Chaos-based random number generators-Part II: Practical realization,”  IEEE Trans. Circ. Syst. I Fundam. Theory Appl. 48(3), 382–385, 2001
\bibitem{callas}John Callas, “Using and Creating Cryptographic-Quality Random Numbers”,  http://www.merrymeet.com/jon/usingrandom.html, 3 June 1996
\bibitem{tim} Tim Matthews,“Suggestions for random number generation in software”,  RSA Data Security Engineering Report, 15 December 1995 (reprinted in RSA Laboratories’Bulletin No.1,22 January  1996).
\bibitem{crypto1} “Cryptographic Random Numbers”, IEEE P1363 Working Draft, Appendix G, 6 February 1997
\bibitem{klaus}  “Zufallstreffer”, Klaus Schmeh and Dr.Hubert Uebelacker,"Cryptographic Security Architecture: Design and Verification", No.14, 1997
\bibitem{donald}Donald Knuth,  Addison-Wesley, “The Art of Computer Programming: Volume 2, Seminumerical Algorithms”,  1981
\bibitem{paul}Alfred Menezes, Paul van Oorschot, and Scott Vanstone “Handbook of Applied Cryptography”,  CRC Press, 1996
\bibitem{crypto2}  Oded Goldreich “Foundations of Cryptography —Fragments of a Book”, February 1995
\bibitem{kuusela}T. Kuusela "Random Number Generation Using a Chaotic Circuit"J. Nonlinear Sci. Vol. 3: pp. 445--458, 1993
\bibitem{matsumoto} T. Matsumoto, L. O. Chua and S. Tanaka, "Simmplext chaotic nonautonomous circuit", Phys. Rev. A, {\bf 30}, 1155 (1984)
\bibitem{andrew} Andrew Rukhin, Juan Soto, James Nechvatal,"A Statistical Test Suite for Random and Pseudorandom Number Generators for Cryptographic Applications"NIST Special Publication 800-22
\bibitem{marsagila} G. Marsaglia, DIEHARD Statistical Tests: http://www.stat.fsu.edu/pub/diehard/.
\bibitem{good}I. J. Good, “ The serial test for sampling numbers and other tests for randomness,” Proc. Cambridge Philos. Soc.. 47, pp. 276-284,1953
\bibitem{baron} M. Baron and A. L. Rukhin, “Distribution of the Number of Visits For a Random Walk,” Communications in Statistics: Stochastic Models. Vol. 15, pp. 593-597,1999
\bibitem{frank} Frank Spitzer, "Principles of Random Walk", Princeton: Van Nostrand, especially p. 269,1964
\end{thebibliography}
\end{document}